\documentclass[12pt]{article}

\usepackage{amsmath, amssymb, amsthm}
\usepackage{enumerate}
\usepackage{multirow}
\usepackage{color}
\usepackage{graphicx}
\usepackage{natbib}

\newcommand{\blind}{0}

\newcommand{\bx}{\mathbf{x}}
\newcommand{\by}{\mathbf{y}}
\newcommand{\bX}{\mathbf{X}}
\newcommand{\bY}{\mathbf{Y}}
\newcommand{\cN}{\mathcal{N}}
\newcommand{\bmu}{\boldsymbol\mu}

\newcommand{\bP}{\mathbf{P}}
\newcommand{\bE}{\mathbf{E}}
\newcommand{\bV}{\mathbf{Var}}

\newcommand{\bEP}{\mathbf{E_P}}

\newcommand{\bcorP}{\mathbf{corr_P}}
\newcommand{\bPB}{\mathbf{P_B}}
\newcommand{\bEB}{\mathbf{E_B}}
\newcommand{\bVB}{\mathbf{Var_B}}

\newtheorem{theorem}{Theorem}[section]

\newtheorem{lemma}[theorem]{Lemma}
\newtheorem{remark}[theorem]{Remark}
\newtheorem{assumption}{Assumption}[section]

\graphicspath{{image/}  {code/}}

\begin{document} 

\def\spacingset#1{\renewcommand{\baselinestretch}%
{#1}\small\normalsize} \spacingset{1}


\if0\blind
{
  \title{\bf A new graph-based two-sample test for multivariate and object data}
  \author{Hao Chen 
\hspace{.2cm}\\
    Department of Statistics, University of California, Davis\\
    and \\
    Jerome H. Friedman \\
    Department of Statistics, Stanford University}
\date{}
  \maketitle
} \fi

\if1\blind
{
  \bigskip
  \bigskip
  \bigskip
  \begin{center}
    {\LARGE\bf A new graph-based two-sample test for multivariate and object data}
\end{center}
  \medskip
} \fi

\bigskip
\begin{abstract}

Two-sample tests for multivariate data and especially for non-Euclidean data are not well explored.  This paper presents a novel test statistic based on a similarity graph constructed on the pooled observations from the two samples.  It can be applied to multivariate data and non-Euclidean data as long as a dissimilarity measure on the sample space can be defined, which can usually be provided by domain experts.  Existing tests based on a similarity graph lack power either for location or for scale alternatives.  The new test utilizes a common pattern that was overlooked previously, and works for both types of alternatives.  The test exhibits substantial power gains in simulation studies.  Its asymptotic permutation null distribution is derived and shown to work well under finite samples, facilitating its application to large data sets.  The new test is illustrated on two applications: The assessment of covariate balance in a matched observational study, and the comparison of network data under different conditions.

\end{abstract}

\noindent%
\vfill

\newpage 
\spacingset{1.45} 
\section{Introduction}
\label{sec:intro}

Two-sample comparison is a classical problem in Statistics.  As we entering the big data era, this problem is encountering new challenges.  Often, researchers want to combine features of subjects together in one test, resulting in multivariate testing problems.  Nowadays, the number of features can be large and there can be underlying structures among them  \citep{de2011clinical, feigenson2014disorganization}.  Also, more complex data types are receiving increasing attention, such as image data and network data \citep{eagle2009inferring, kossinets2006empirical}.  Effectively comparing samples of these data types is a challenging but important problem.  For parametric approaches, their power decreases quickly as the dimension increases unless strong assumptions are made to facilitate the estimation of the large number of parameters, such as the covariance matrix.  In this work, we propose and study a nonparametric testing procedure that works for both multivariate data and object data against general alternatives.

Nonparametric testing for two sample differences has a long history and rich literature.  Some well known examples include the Kolmogorov-Smirnov test, the Wilcoxon test, and the Wald-Wolfowitz runs test (see \cite{gibbons2011nonparametric} for a survey).  People have tried to generalize these procedures to multidimensional settings from long time ago.  \cite{weiss1960two} generalized the Wald-Wolfowitz runs test through drawing the largest possible non-overlapping spheres around one sample and count the number of spheres that do not contain observations from the other sample.  However, the null distribution of the statistic is not known and is distribution dependent.  \cite{darling1957kolmogorov} and \cite{bickel1969distribution} generalized the Kolmogorov-Smirnov test using the multivariate empirical cdf, while in order for these methods to work well, the required sample size is exponential in dimension.

 \cite{friedman1979multivariate} proposed the first practical test that can be applied to data with arbitrary dimension.  They used the pairwise distances among the pooled observations to construct a minimum spanning tree (MST), which is a spanning tree that connects all observations with the sum of distances of edges in the tree minimized.  Tests were conducted based on the MST.  The principal one is a count statistic on the number of edges that connect nodes (observations) from different samples, which can be viewed as a generalization of the Wald-Wolfowitz runs test to the multidimensional setting.   The rationale of the test is that, if the two samples are from different distributions, observations would be preferentially closer to others from the same sample than those from the other sample.  Thus edges in the MST would be more likely to connect observations from the same sample.  The test rejects the null if the number of between-sample edges is significantly \emph{less} than what is expected.   We call this test the \emph{edge-count test} for easy reference. 

The edge-count test can be applied to other similarity graphs.  \cite{friedman1979multivariate} also considered denser graphs, e.g., the $k$-MST, which is the union of the 1st, \dots, $k$th MSTs, where a $k$th MST is a spanning tree connecting all observations that minimizes the sum of distances across edges subject to the constraint that this spanning tree does not contain any edge in the 1st, \dots, $k\text{-}1$th MST(s).  They showed that the edge-count test on a 3-MST is usually more powerful than the edge-count test on a 1-MST.  \cite{schilling1986multivariate} and \cite{henze1988multivariate} used $k$-nearest neighbor ($k$-NN) graphs where each observation is connected to its $k$ closest neighbors.  \cite{rosenbaum2005exact} proposed to use the minimum distance non-bipartite pairing (MDP).  This divides the $N$ observations into $N/2$ (assuming $N$ is even) non-overlapping pairs in such a way as to minimize the sum of $N/2$ distances within pairs.  For an odd $N$, Rosenbaum suggested creating a pseudo data point that has distance 0 to all observations, and later discarding the pair containing this pseudo point.  The edge-count test on the MDP is exactly distribution free because the structure of the MDP only depends on the sample size under the null hypothesis.  

\cite{friedman1979multivariate} proposed other tests based on the MST as well.  They viewed the MST as a generalization of the ``sorted list'' and formed generalizations of the Smirnov test and the radial Smirnov test.  They also proposed a degree test on the MST by pooling observations into a $2\times 2$ contingency table according to (i) whether the observation is from the sample $\bX$ or not, and (ii) whether the observation has degree 1 in the MST or not, and tested their independence.  The generalizations of the Smirnov test and the radial Smirnov test in \cite{friedman1979multivariate} required the graph being a tree; while the degree test can easily be generalized to other types of graphs.  \cite{rosenbaum2005exact} also proposed another test based on the MDP by using the rank of the distance within the pairs, which is thus restricted to MDPs.

%

All these tests based on a similarity graph on observations can be applied to non-Euclidean data as long as a similarity measure on the sample space can be defined.  \cite{maa1996reducing} provided the theoretical basis for this type of test for the multivariate setting. They showed that, under mild conditions, two multivariate distributions are equivalent if and only if the distributions of interpoint distances within each distribution and between the distributions are all equivalent.   
In addition, \cite{henze1999multivariate} showed that the edge-count test on the MST is consistent against all alternatives for multivariate distributions when the MST is constructed using the Euclidean distance.  Since non-Euclidean data can often be embedded into a high-dimensional Euclidean space, these results also provide us with confidence in applying these tests to non-Euclidean data.

However, the reality is not as promising for these existing tests.  The two basic types of alternatives are location and scale alternatives.  Although all these tests were proposed for general alternatives, none of them is sensitive to both kinds of alternatives in practical settings.  Asymptotically, the edge-count test is able to distinguish both types of alternatives as proved by \cite{henze1999multivariate}.  In practice, the edge-count test has low or even no power for scale alternatives when the dimension is moderate to high unless the sample size is astronomical due to the curse-of-dimensionality.  The detailed reason is given in Section \ref{sec:problem}. 
For the other tests mentioned above, \cite{friedman1979multivariate} showed that the generalization of the Smirnov test has no power for scale-only alternatives and the generalization of the radial Smirnov test and the degree test on the MST have no power for location-only alternatives.  The rank test on the MDP proposed by \cite{rosenbaum2005exact} has similar rationale and performance to the edge-count test on the MDP. 

To solve the problem, we propose a new test which utilizes a common pattern in both types of alternatives and thus works well for them both and even general location-scale alternatives.   The details of the new test are discussed in Section \ref{sec:statistic}.   We study the power of the proposed test under different scenarios and compare it to other existing tests in Section \ref{sec:power}.  In Section \ref{sec:asymptotic}, we derive the asymptotic permutation null distribution of the test statistic and show how the $p$-value approximation based on the asymptotic null distribution works for finite samples.  In Section \ref{sec:application}, the proposed test is illustrated by two applications: Appraising covariate balance in matched college students, and comparing phone-call networks under different conditions.  We discuss a few other test statistics along the same line in Section \ref{sec:discussion}, and conclude in Section \ref{sec:conclusion}.

\section{The problem}
\label{sec:problem}


Although \cite{henze1999multivariate} proved that the edge-count test on MST constructed on Euclidean distance is consistent against all alternatives, the test does not work under some common scenarios.
As an illustration example, we consider the testing of two samples, both sample sizes are 1,000, from two distributions, $F_\bX = \cN(\mathbf{0}, I_{d})$ and $F_\bY =\cN(\bmu, \sigma^2 I_{d})$, $d=100$, respectively, where $I_d$ is a $d\times d$ identity matrix.  Here, $\cN(\bmu, \Sigma)$ denotes a multivariate normal distribution with mean $\bmu$ and covariance matrix $\Sigma$.  We use the common notation $\|\cdot\|_2$ to denote $L_2$ norm.
\begin{itemize}
\item Scenario 1: Only mean differs, $\|\bmu\|_2 = 1$, $\sigma=1$.
\item Scenario 2: Both mean and variance differ, $\|\bmu\|_2=1$, $\sigma=1.1$.
\end{itemize}
In most simulation runs, the edge-count test on MST constructed on the Euclidean distance rejects the null hypothesis under scenario 1, but does not reject the null hypothesis under scenario 2.  Of course, the additional difference in variance in scenario 2 would not make the two distributions more similar.  So what happened here?

We study a typical simulation run under scenario 2.  The MST constructed on the 2,000 points (observations) based on the Euclidean distance contains 979 between-sample edges, which is quite close to its null expectation (1,000).  Thus, the edge-count test does not reject the null hypothesis.  However, if we take a closer look, 
 in the MST, there are 991 edges connecting points within sample $\bX$, but only 29 edges connecting points within sample $\bY$.  The fact that almost all points from sample $\bY$  find points from sample $\bX$ closer contributes a lot to the between-sample edges, making the edge-count test have low power under this scenario.  Then why do points in sample $\bY$ find points in sample $\bX$ closer?  In Figure 2, we show the boxplots of distances of points in each sample to the center of all points from both samples.  We see that the two samples are well separated into two layers: Sample $\bX$ in the inner layer and sample $\bY$ in the outer layer.


\begin{figure}[!htp]\label{fig:distance}
\centering
\includegraphics[width=.6\textwidth]{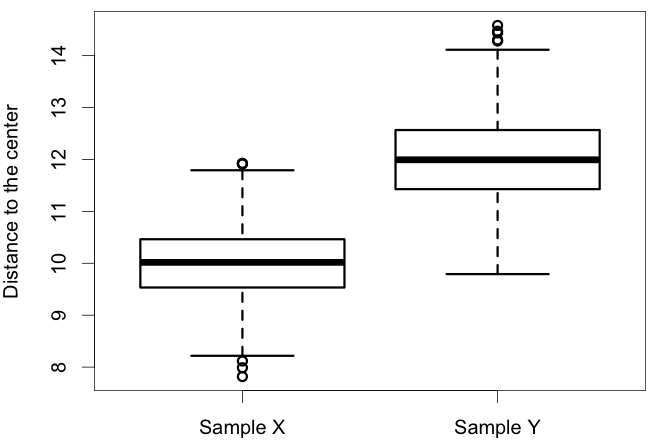}
\caption{Boxplots of the distances of the points in each sample to the center of all points from both samples from a typical simulation run under scenario 2.}
\end{figure}

When the dimension is moderate to high and the two distributions differ in scale, the phenomenon that points in the outer layer find themselves to be closer to points in the inner layer than other points in the outer layer is common unless the number of points in the outer layer is extremely large. 
 The reason is that the volume of a $d$-dimensional space increases exponentially in $d$.  When $d$ is large, we can put a huge number of points on the unit surface such that no pair of them is closer than 1.  Then, each point on the unit surface would find the \emph{origin} to be closer than any other point on the unit surface.  If there are points on an inner layer inside of the unit surface, then most of the points on the unit surface would find points in that inner layer to be closer than their closest points on the unit surface.  This argument can be extended to any pair of distributions differing in scale under moderate to high dimension.   

\begin{figure}[!htp]
  \centering
  \includegraphics[width=.6\textwidth]{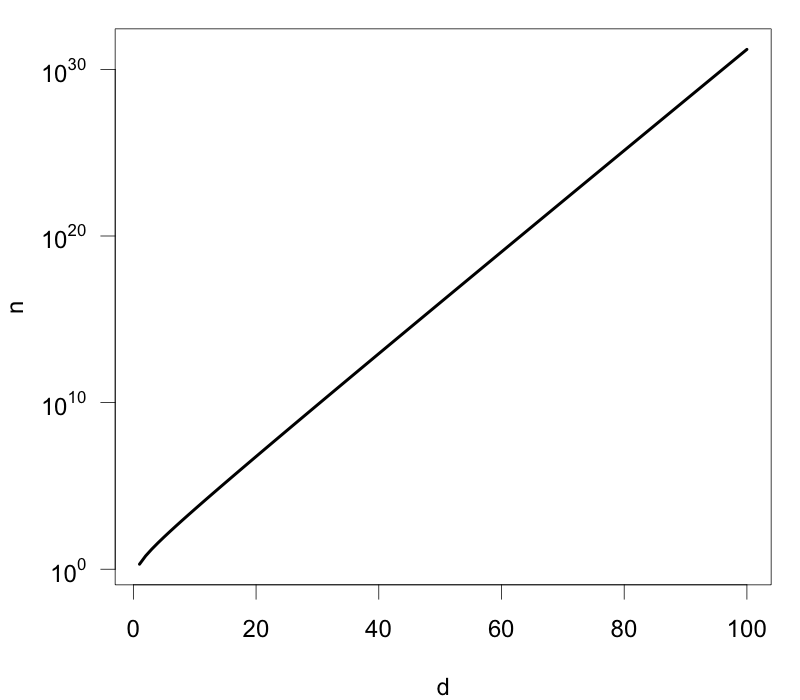}
  \caption{The approximate number of points that can be put on the $d$-dimensional unit ball such that the distance between any two points is larger than 1.  The $x$-axis is the dimension and the $y$-axis (in logarithmic scale) is the approximate number of points.}
  \label{fig:dvN}
\end{figure}

To give an idea on how large the number can be, we approximate it by the number of non-overlapping ($d-1$)-dimensional balls with radius $0.5$ on the surface of the $d$-dimensional unit ball, which can further be approximated by the ratio of the surface area of the $d$-dimensional unit ball,
$$\frac{d\pi^{d/2}}{\Gamma(d/2+1)}, $$
over the volume of the ($d-1$)-dimensional ball with radius $1/2$,
$$\frac{\pi^{(d-1)/2} (1/2)^{d-1}}{\Gamma((d-1)/2+1)}.$$
This gives 
$$\frac{\sqrt{\pi} d \Gamma(d/2+1/2)}{\Gamma(d/2+1)} \times 2^{d-1}.$$
This approximate number is plotted versus dimension ($d$) in Figure \ref{fig:dvN}.  We can see that the number is exponential in $d$ (the $y$-axis is in a logarithmic scale).   When the dimension is 30, the number is around $10^{10}$.  When the dimension is 65, the number is about $10^{20}$.  These numbers can hardly be achieved in reality in terms of the number of observations in one sample.  Therefore, in practice, the edge-count test on a similarity graph that connects observations ``closer'' in the usual sense, e.g., Euclidean distance, does not work under the scale alternative when the dimension is moderate to high. 

To solve this problem, there are some options.  One way is to define a new sense of ``closeness''.  For example, if we know the change is in scale and the distribution is isotropic, we can define the closeness based on the distance to the center of all points:  Points are closer if their distances to the center are more similar.  However, this relies heavily on the type of the alternative and the ``closeness'' that works well for one alternative can work poorly for another. 

In this paper, we adopt a different approach.  We construct the similarity graph in the usual sense of ``closeness" but define a new test statistic that utilizes a common pattern in the two types of alternatives.  The details are in the following section.



\section{A new test statistic}
\label{sec:statistic}

There is a key fact: In either location or scale alternatives, in the similarity graph constructed through the usual sense of ``closeness", the numbers of within-sample edges for the two samples deviate from their null expectations, though the direction of deviations can be different.  In location alternatives, both numbers of within-sample edges for the two samples would be more than their null expectations, so the edge-count test works.  In scale alternatives and when the dimension is moderate to high, the number of within-sample edges for the sample in the inner layer would be more than its null expectation, while the number of within-sample edges for the sample in the outer layer would be less than its null expectation, making the edge-count test have low or no power.  We can, however, incorporate both directions of deviations together and construct a test statistic that is powerful for both types of alternatives. 

Before defining the test statistic, we first give the formal formulation of the problem.  We have two independent samples $\{\bX_1,\dots,\bX_n\}$ and $\{\bY_1,\dots,\bY_m\}$, with the formal independent and identically distributed according to a distribution $F_\bX$ and the latter $F_\bY$. 
We test for $H_0: F_\bX = F_\bY$ versus a general alternative $H_A: F_\bX\neq F_\bY.$  We let $N=n+m$ be the total sample size.

Under the null hypothesis $F_\bX=F_\bY$, the group identity is exchangeable.  In the following, we work under the permutation null distribution, which places $1/\binom{N}{n}$ probability on each of the $\binom{N}{n}$ choices of $n$ out of the total $N$ observations as the $\bX$-sample.  When there is no further specification, we denote by $\bP$, $\bE$, $\bV$ probability, expectation, and variance, respectively, under the permutation null distribution. 

The new test statistic we propose utilizes a similarity graph constructed on the pooled observations.  Let $G$ be an undirected similarity graph constructed in terms of usual ``closeness'' on the observations, such as a MST constructed using $L_2$ or $L_1$ distance.  We restrict $G$ to have no multi-edge.  That is, any pair of nodes is connected by at most one edge.  The $k$-MST by construction satisfy this restriction.  The similarity graph need not be derived from a similarity measure.  It can be directly provided by domain experts based on domain knowledge. 

We use $G$ to refer to both the graph and its set of edges, when the vertex set is implicitly obvious.  The symbol $|\cdot |$ is used to denote the size of a set, so $|G|$ is the number edges in $G$.  For any event $x$, we let $I_x$ be the indicator function that takes value 1 if $x$ is true, and 0 otherwise.  We pool observations and index them by $1,\dots,N$.  Let $g_i=0$ if the observation is from sample $\bX$ and 1 otherwise.  For an edge $e=(i,j)$, we define
\begin{align} 
  J_e & = \left\{ \begin{array}{lll} 0 & \text{if} & g_i\neq g_j  \\ 1 & \text{if} & g_i=g_j=0  \\ 2 & \text{if} & g_i=g_j=1
\end{array} \right., \nonumber \\
  R_k & = \sum_{e\in G} I_{J_e=k}, \quad k=0,1,2. \label{eq:Ri}
\end{align}
Then $R_0$ is the number of between-sample edges (which is the test statistic for the edge-count test), $R_1$ is the number of edges connecting observations both from sample $\bX$, and $R_2$ is the number of edges connecting observations both from sample $\bY$.

The new test statistic is defined as follows:
$$S=(R_1 - \mu_1, R_2 - \mu_2) \Sigma^{-1} \left(\begin{array}{c} R_1 - \mu_1 \\ R_2 - \mu_2 \end{array} \right),$$
where $\mu_1 = \bE(R_1), \mu_2 = \bE(R_2)$, and $\Sigma$ is the covariance matrix of the vector $(R_1,R_2)^\prime$ under the permutation null distribution.  The test statistic is defined in this way so that either direction of deviations of the number of within-sample edges from its null expectation would contribute to the test statistic.  Under the location-alternative, or the scale-alternative for low-dimensional data, we would expect both $R_1$ and $R_2$ to be larger than their null expectations, then $S$ would be large.  Under the scale-alternative for moderate/high-dimensional data, the number of within-sample edges for the sample with a smaller variance is expected to be larger than its null expectation, and the number of within-sample edges for the sample with a larger variance is expected to be smaller than its null expectation, then $S$ would also be large.  Therefore, the test defined in this way is sensitive to both location and scale alternatives.

The analytic expressions for $\mu_1, \mu_2$, and $\Sigma=(\Sigma_{i,j})_{i,j=1,2}$ can be calculated through combinatorial analysis.  They are given in the following lemma.
\begin{lemma}\label{lemma:EV}
We have
  \begin{align*}
    \mu_1 &  = |G| \frac{n(n-1)}{N(N-1)} \\
    \mu_2 &  = |G| \frac{m(m-1)}{N(N-1)}  \\
    \Sigma_{11} & = \mu_1(1-\mu_1) + 2C \frac{n(n-1)(n-2)}{N(N-1)(N-2)}  \\ 
    & \quad + \left(|G|(|G|-1) - 2C\right) \frac{n(n-1)(n-2)(n-3)}{N(N-1)(N-2)(N-3)} \\
    \Sigma_{22} & = \mu_2(1-\mu_2) + 2C \frac{m(m-1)(m-2)}{N(N-1)(N-2)} \\ 
    & \quad + \left(|G|(|G|-1) -2C\right)\frac{m(m-1)(m-2)(m-3)}{N(N-1)(N-2)(N-3)} \\
    \Sigma_{12} & = \Sigma_{21}  = \left(|G|(|G|-1) -2C\right)\frac{nm(n-1)(m-1)}{N(N-1)(N-2)(N-3)} - \mu_1\mu_2.
  \end{align*}
where $C=\frac{1}{2}\sum_{i=1}^N|G_i|^2 - |G|$, with $G_i$ being the subgraph in $G$ that includes all edge(s) that connect to node $i$.
\end{lemma}

The quantity $C$ is the number of edge pairs that share a common node.  The proof to this lemma is in Appendix \ref{sec:proof-lemma-EV}. 

To ensure that the proposed test statistic is well defined, $\Sigma$ needs to be invertible.  

\begin{theorem}\label{thm:invertible}
For $N>3$, $|G|>0$, the proposed  test statistic $S$ is well defined except for the following two kinds of graphs.
\begin{enumerate}[\quad (1)]
\item All nodes have exactly the same degree, i.e., $|G_1|=|G_2|=\dots =|G_N|$.
\item The perfect star-shaped graph, that is, one node has degree $N-1$ and all other nodes have degree 1.
\end{enumerate}
\end{theorem}

\begin{proof}
The elements in $\Sigma$ can be re-organized as
\begin{align*}
\Sigma_{11} & = \frac{nm(n-1)(m-1)}{N(N-1)(N-2)(N-3)}\left(|G| + \frac{n-2}{m-1}\left(\sum_{i=1}^N|G_i|^2 - \frac{4|G|^2}{N} \right) -\frac{2}{N(N-1)}|G|^2 \right), \\
\Sigma_{12} & = \frac{nm(n-1)(m-1)}{N(N-1)(N-2)(N-3)}\left(|G| -\left(\sum_{i=1}^N|G_i|^2 - \frac{4|G|^2}{N} \right) -\frac{2}{N(N-1)}|G|^2 \right), \\
\Sigma_{22} & =  \frac{nm(n-1)(m-1)}{N(N-1)(N-2)(N-3)}\left(|G| + \frac{m-2}{n-1}\left(\sum_{i=1}^N|G_i|^2 - \frac{4|G|^2}{N} \right) -\frac{2}{N(N-1)}|G|^2 \right),
\end{align*}
and the determinant of $\Sigma$ can be expressed as
\begin{align*}
|\Sigma|& = \frac{nm}{N(N-1)(N-2)}\left(\sum_{i=1}^N|G_i|^2 - \frac{4|G|^2}{N} \right)\left((N-2)|G|+\frac{2}{N-1}|G|^2- \sum_{i=1}^N|G_i|^2\right).
\end{align*}

From the Cauchy-Schwarz inequality, we know that 
$$\sum_{i=1}^N|G_i|^2\geq \frac{4|G|^2}{N}.$$
Here, equality only holds when $|G_i|$'s are equal for all $i$'s, which leads to the first kind of graph for which the $\Sigma$ is non-invertible.

We next figure out the kind of graph that $(N-2)|G|+\frac{2}{N-1}|G|^2- \sum_{i=1}^N|G_i|^2=0$.

If $|G|<N-1$, then $\sum_{i=1}^N|G_i|^2 \leq |G|^2+|G|.$
So 
\begin{align*}
(N-2)|G|+\frac{2}{N-1}|G|^2- \sum_{i=1}^N|G_i|^2 &  \geq (N-2)|G|+\frac{2}{N-1}|G|^2-|G|^2-|G|  \\
& = (N-3)|G|\left(1-\frac{|G|}{N-1}\right)>0.
\end{align*}

 If $|G|\geq N-1$, we let $|G|=\sum_{i=1}^s(N-i)+t$ where $1\leq s\leq N-1$ and $0\leq t< N-s$. 
Then the graph with the maximum $\sum_{i=1}^N |G_i|^2$ is that $s$ node(s) connect to every other node (side node) and one side node connects to $t$ other side nodes.  Therefore,
\begin{align*}
\sum_{i=1}^N |G_i|^2 & \leq s(N-1)^2 + (s+t)^2 + t(s+1)^2 + (N-s-1-t)s^2 \\
& = s(N-1)^2 + s^2(N-1)-s^2(s-1)+4st+t^2+t.
\end{align*}
Noticing that $|G|=\sum_{i=1}^s(N-i)+t = s(N-1)-s(s-1)/2 + t$, we have,
\begin{align*}
(N-2)|G|&+\frac{2}{N-1}|G|^2- \sum_{i=1}^N|G_i|^2 \\
& \geq -\frac{N-3}{N-1}t^2 + \left(N-3-\frac{2s(s-1)}{N-1} \right)t + \frac{s(s-1)(N-s)(N-s-1)}{2(N-1)}:=h(N,s,t).
\end{align*}
Now, $h(N,s,t)$, $0\leq t\leq N-s-1$, is a quadratic function of $t$ and the coefficient of $t^2$ is negative, so its minimum is achieved at either $t=0$ or $t=N-s-1$.  It follows that
\begin{align*}
h(N,s,0) & = \frac{s(s-1)(N-s)(N-s-1)}{2(N-1)}\geq 0, \\
h(N,s,N-s-1) & = (N-s-1)\left(-\frac{N-3}{N-1}(N-s-1) + N-3-\frac{2s(s-1)}{N-1} + \frac{s(s-1)(N-s)}{2(N-1)} \right)  \\
& = \frac{s(s+1)(N-1-s)(N-2-s)}{2(N-1)}\geq 0.
\end{align*}
Therefore, when $|G|\geq N-1$, we have $(N-2)|G|+\frac{2}{N-1}|G|^2- \sum_{i=1}^N|G_i|^2 \geq 0.$  Here, equality holds only when (i) $t=0, s=1$ and the graph is perfectly star-shaped; or (ii) $t=0, s=N-1$ and $G$ is the complete graph (all nodes have degree $N-1$).

\end{proof}

\begin{remark}\label{remark:Sigma}
If $N$ is even, and a $k$-MDP is constructed, then all nodes have degree $k$ and $\Sigma$ is non-invertible.  It would not be a problem when $N$ is odd (and $k<N$).  However, since $\sum_{i=1}^N|G_i|^2-4|G|^2/N$  is very small in $k$-MDP compared to $|G|$, the condition number of $\Sigma$ is large and its inversion is unstable.  The same problem arises when the graph is roughly star-shaped.  Therefore, if such similarity graphs are obtained, we recommend not to use the proposed test statistic or one can seek better ways to construct the graph. 
\end{remark}

In the following, we refer to $|G|,n,m\rightarrow\infty,\ n/(n+m)\rightarrow p\in(0,1)$ as the \emph{\textbf{usual limiting regime}}.

\begin{remark}\label{remark:Sigmalimit}  
In the usual limiting regime,
when $\sum_{i=1}^N|G_i|^2 - \frac{4|G|^2}{N}=O(|G|)$, which is commonly achieved for the $k$-MST, $k=O(1)$, (see the proof of Theorem \ref{thm:asym-kMST}), the limiting quantities for $\mu_1/|G|$, $\mu_2/|G|$ and $\Sigma/|G|$ are
\begin{align*}
\lim_{N\rightarrow\infty} \frac{\mu_1}{|G|} & = p^2, \\
\lim_{N\rightarrow\infty} \frac{\mu_2}{|G|} & = q^2, \\
\lim_{N\rightarrow\infty} \frac{\Sigma}{|G|} & = p^2 q^2 \left( \begin{array}{cc}1+rp/q & 1-r \\ 1-r & 1+ r q/p  \end{array} \right),
\end{align*}
where $q=1-p$, and $r=\lim_{N\rightarrow\infty}\sum_{i=1}^N(|G_i|^2 - 4|G|^2/N)/|G|$.  

If the graph is a $k$-MST, $k=O(1)$, then $|G|=k(N-1)$, and $r$ is a function of $k$ and the dimension of the data if they lie in an Euclidean space and the $k$-MST is constructed based on the Euclidean distance (see the proof of Theorem \ref{thm:asym-kMST}).
\end{remark}

The topology of $G$ completely determines the permutation distribution of the test statistic.  One can compute higher moments in the same manner as the variance in Lemma \ref{lemma:EV}, which is however very tedious when the order of moments is high.  To obtain the permutation $p$-value, for small enough sample size, it is feasible to calculate directly the distribution of $S$ over all permutations.  This, however, can be time consuming for large sample sizes.  We show that the permutation null distribution of $S$ approaches the $\chi^2_2$ distribution under some mild conditions on the graph (see details in Section \ref{sec:asymptotic}).

\section{Power comparison}
\label{sec:power}
The utility of the test presented in the previous section lies in its power to discriminate against a wide variety of alternative hypotheses.  In this section, we present results of various simulation studies in examining the power of the test for several alternative hypotheses in various dimensions.

To have a baseline for comparison, we choose the distribution to be multivariate Gaussian distribution so that we have the asymptotically most powerful tests based on the normal theory --  the Hotelling's two-sample $T^2$ test if assuming equal covariance matrices [``Hotelling's $T^2$"], and the generalized likelihood ratio test if not assuming equal covariance matrices [``GLR''].  The test statistic of the Hotelling $T^2$ is 
$$\frac{nm}{N} (\bar{\bx}-\bar{\by})^\prime W^{-1} (\bar{\bx}-\bar{\by}),$$ $$\text{ with } \bar{\bx}=\frac{\sum_{i=1}^n \bx_i}{n},\ \bar{\by}=\frac{\sum_{i=1}^m \by_i}{m},\ W=\frac{\sum_{i=1}^n (\bx_i-\bar{\bx})(\bx_i-\bar{\bx})^\prime + \sum_{i=1}^m (\by_i-\bar{\by})(\by_i-\bar{\by})^\prime}{N-2},$$ 
and the test statistic of GLR is 
$N\log|\hat{\Sigma}_0| - n \log|\hat{\Sigma}_\bx| - m\log|\hat{\Sigma}_\by|,$ where $\hat{\Sigma}_0$, $\hat{\Sigma}_\bx$, and $\hat{\Sigma}_\by$ are the maximum likelihood estimators of the covariance matrix of the whole data, sample $\bX$ and sample $\bY$.

In addition to the two tests based on the normal theory, we include in the comparison the new test on the MST, 3-MST and 5-MST [``$S$: 1-,3-,5-MST''], the edge-count test on MSTs [``$R_0$: 1-,3-,5-MST''] and on MDPs [``$R_0$: 1-,3-,5-MDP''], as well as the degree test on the MST proposed in \cite{friedman1979multivariate} [``deg 1''].  All MSTs and MDPs are constructed using the Euclidean distance. 

Table \ref{tab:power} shows results for two multivariate Gaussian distributions with different means (the $L_2$ distance of the two means is $\Delta$).  The results are from low dimension ($d=2$) to high dimension ($d=100$).   For each case, the specific alternative hypothesis was chosen so that the tests have moderate power.  We see that Hotelling's $T^2$ test is doing very well when the dimension is low to moderate since all assumptions for Hotelling's $T^2$ test hold.  However, when the dimension becomes higher, the power of Hotelling's $T^2$ test is outperformed by the edge-count tests and the new test.  In the table, we show only up to $d=100$, while the edge-count tests and the new test are not limited by the dimension.  Based on the current trend, even if we keep the same amount of $\Delta$, the power of the edge-count tests and the new test decrease slowly as the dimension increases.  This is the scenario where the edge-count test works and we see that the new test is only slightly worse than the edge-count test.


%
 
\begin{table}[!htp]
 \caption{\label{tab:power} Number of trials (out of 100) with significance less than 5\%,  normal data.  The means of the two distributions differ in $\Delta$ in $L_2$ distance.  $n=m=50$.}
  \centering 
  \begin{tabular}{ccccccccc}
    \hline
    \multicolumn{8}{c}{Location alternatives} \\
    $d$ & 2 & 10 & 30 & 50 & 70 & 90 & 100 \\ 
    $\Delta$ & 0.6 & 0.8 & 1.1 & 1.4 & 1.7 & 2 & 2 \\ \hline
    Hotelling's $T^2$ & \textbf{77} & \textbf{71} & \textbf{74} & \textbf{76} & 70 & 26 & - \\
    GLR & 52 & 30 & 14 &  - & - & - & - \\
    $R_0$: 1-,3-,5-MST & 22 35 40 & 12 35 47 & 27 46 49 & 37 67 73 & 41 76 \textbf{89} & 61 85 \textbf{92} & 57 85 \textbf{90} \\
    $R_0$: 1-,3-,5-MDP & 9 25 32 & 10 26 38 & 18 36 43 & 21 47 64 & 27 63 86 & 41 74 89 & 50 75 87  \\
    deg 1 & 4 & 6 & 4 & 4 & 3 & 4 & 4 \\ 
    $S$: 1-,3-,5-MST & 10 22 24 & 9 23 34 & 20 30 34 & 25 40 59 & 23 54 80 & 36 76 83 & 34 74 82  \\ \hline
  \end{tabular}
\end{table}

Table \ref{tab:power2} shows results for two multivariate Gaussian distributions with different variances (differ in a multiple of $\sigma$).  Since the equal covariance matrices assumption for Hotelling's $T^2$ test does not hold here, Hotelling's $T^2$ test is doing poorly.  The GLR test is doing well in very low dimension ($d=2$).  When the dimension increases a bit ($d=5$), it is already outperformed by the new test.  The reason is that the number of parameters that need to be estimated for the GLR test increases quickly as the dimension increases and its power decreases quickly, while the new test is relatively dimension-free.  This is the scenario where the edge-count test becomes not working properly as the dimension increases.  We see that the edge-count test is working okay in low dimensions, but has much lower power than the new test as the dimension increases.  The degree test, which has no power in the location-only alternative (Table \ref{tab:power}), is powerful here, but it is dominated by the new test.

\begin{table}[!htp]
  \caption{\label{tab:power2}  Number of trials (out of 100) with significance less than 5\%, normal data, $n=m=50$.}
  \centering
  \begin{tabular}{ccccc}
    \hline
    \multicolumn{5}{c}{Scale alternatives} \\
$d$ & 2 & 5 & 10 & 20 \\
$\sigma$ & 1.4 & 1.25 & 1.2 & 1.15 \\ \hline
Hotelling's $T^2$ & 7 & 7  & 5 & 5 \\ 
GLR & \textbf{69} & 42 & 28 & 12 \\
$R_0$: 1-,3-,5-MST & 22 34 41 & 12 22 24 & 7 17 28 & 7 15 18 \\
$R_0$: 1-,3-,5-MDP & 16 28 36 & 12 14 17 & 7 9 18 & 5 5 10 \\
deg 1 & 8 & 27 & 59 & 62 \\
$S$: 1-,3-,5-MST & 20 43 {56} & 37 \textbf{64} \textbf{64} & 57 76 \textbf{78} & 66 73 \textbf{80} \\ \hline
  \end{tabular}  
\end{table}

We also compare all the tests for log normal data.  The distributions are products of independent log normal distributions with alternatives differing in the location parameter (the difference of the two location parameters is $\Delta$).  Changing location parameter changes both the mean and variance of a log normal distribution, so the alternative is both location and scale.  We see that, when the dimension is moderate to high, the new test dominates all other tests (Table \ref{tab:power3}).

\begin{table}[!htp]
  \caption{\label{tab:power3}  Number of trials (out of 100) with significance less than 5\%, product log normal data, $n=m=50$.}
  \centering
  \begin{tabular}{ccccccc}
    \hline
    \multicolumn{7}{c}{Log location alternatives} \\
$d$ & 2 & 10 & 30 & 50 & 70 & 90 \\
$\Delta$ & 0.8 & 1 & 1.3 & 1.3 & 1.5 & 1.7 \\ \hline
Hotelling's $T^2$ & \textbf{82} & \textbf{81} & \textbf{79} & 52 & 39 & 20 \\
GLR & 27 & 18 & 16 & - & - & - \\
$R_0$: 1-,3-,5-MST & 38 58 62 & 26 49 58 & 22 45 51 & 14 44 52 & 16 48 60 & 21 42 53 \\
$R_0$: 1-,3-,5-MDP & 25 44 54 & 18 34 50 & 11 31 40 & 11 23 35 & 15 36 49 & 12 34 47 \\
deg 1 & 4 & 10 & 29 & 41 & 50 & 47 \\
$S$: 1-,3-,5-MST & 19 39 53 & 25 46 57 & 43 52 61 & 40 57 \textbf{62} & 46 65 \textbf{69} & 51 69 \textbf{75} \\ \hline
  \end{tabular}  
\end{table}




From the simulation results, the new test exhibits high power for both location and scale alternatives, as well as for general location-scale alternatives.  Unless we are very confident that the alternative is location-only, the new test is preferred in moderate to high dimensions. 

\begin{remark}
In all simulation studies, the power of both the edge-count test and the new test increase when the similarity graph becomes denser, from a 1-MST to a 5-MST.  This is reasonable because 5-MST has more ``similarity'' information than 1-MST does.  To the other extreme, if we make the similarity graph too dense, we would include edges that do not provide any ``similarity'' information or even provide counter information.  This would reduce the power of the test.  For example, the test on the complete graph would have no power at all.  Therefore, there is an optimal density of the graph for each application.   For the simulation settings, the 5-MST has not achieved the optimal point since the trend of increasing power from a 1-MST to a 5-MST has not been stabilized.  On the other hand, if we make the graph denser, the computation cost is also higher.  These tradeoffs are not explored in this paper.  From a practical point of view, the 5-MST is a reasonable initial choice when the sample sizes are in hundreds.
\end{remark}

\begin{figure}[!htp]
  \centering
Edge-count test ($R_0$) \quad \quad \quad \quad \quad  \quad \quad \quad \quad The new test ($S$)\\
\vspace{0.5em}

  \includegraphics[width=.4\textwidth]{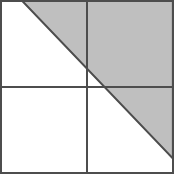} \quad \quad
  \includegraphics[width=.4\textwidth]{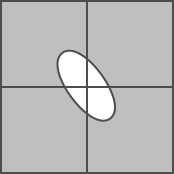}
  \caption{Rejection regions (shaded) of the edge-count test ($R_0$) and the new test ($S$).  The horizontal and vertical axises are $R_1-\bE(R_1)$ and $R_2-\bE(R_2)$, respectively.}
  \label{fig:cartoon}
\end{figure}

To have a better understanding of the edge-count test and the new test, we plot their rejection regions in Figure \ref{fig:cartoon}.  The horizontal and vertical axes in both plots are $R_1-\bE(R_1)$ and $R_2-\bE(R_2)$, respectively.  When there is only a locational difference or the dimension is very low, the alternative appears in the first quadrant, so the edge-count test has a slightly higher power than the new test.  But the new test can gain power quickly as the amount of change increases.  When there is a scale change and the dimension is moderate to high, the alternative would appear in the second or the fourth quadrant unless the sample size is astronomically large.  The new test still has good power, while the edge-count test has very poor power.  The consistency of the new test statistic under a multivariate setting is discussed in Section \ref{sec:consistency}.

%

\section{Asymptotics}
\label{sec:asymptotic}
When the sample size is small, we can obtain the permutation $p$-value directly from the permutation distribution of $S$.  This is time consuming when the sample size is large.  In this section, we show that, in the usual limiting regime (see its definition in Remark \ref{remark:Sigmalimit}), the permutation null distribution of $S$ approaches the $\chi^2_2$ distribution under some mild conditions on the similarity graph $G$.  This facilitates the application of the new test to large data sets.  If the data is multivariate, then the $k$-MST, $k=O(1)$, based on the Euclidean distance satisfies all the conditions required for getting the asymptotic null distribution.  In addition, if the two multivariate distribution are continuous and differ on a set of positive measure, then the proposed test based on the $k$-MST, $k=O(1)$, is consistent against all alternatives.  We also study how well the asymptotic null distribution works in approximating $p$-values for finite samples.   

\subsection{Asymptotic null distribution}

Before stating the theorem, we define two additional terms on the similarity graph $G$:
 \begin{align}
  A_e & = \{e\} \cup \{e^\prime\in G: e^\prime \text{ and } e \text{ share a node}\},\nonumber \\
  B_e & = A_e \cup \{e^{\prime\prime}\in G: \exists \  e^\prime\in A_e, \text{ such that } e^{\prime\prime} \text{ and } e^\prime \text{ share a node}\}.\nonumber
\end{align}
So $A_e$ is the subgraph in $G$ that connects to edge $e$, and $B_e$ is the subgraph in $G$ that connects to any edge in $A_e$.


\begin{theorem}\label{thm:asym}
  If $|G|=O(N),\sum_{i=1}^N|G_i|^2=O(N)$, $\sum_{e\in G} |A_e||B_e| = o(N^{1.5})$, and  $\sum_{i=1}^N|G_i|^2-4|G|^2/N=O(N)$, in the usual limiting regime, under the permutation null, 
  \begin{equation}
    \label{eq:asym}
    S := (R_1 - \mu_1, R_2 - \mu_2) \Sigma^{-1} \left(\begin{array}{c} R_1 - \mu_1 \\ R_2 - \mu_2 \end{array} \right)\overset{\mathcal{D}}{\rightarrow} \chi_2^2.
  \end{equation}
\end{theorem}

This theorem can be proved through extensions of the methods used in \cite{chen2013graph} and \cite{chen2015graph}.  The complete proof is in Appendix \ref{sec:proof-thm-asym}.   

The condition $\sum_{i=1}^N|G_i|^2-4|G|^2/N =O(N)$ ensures the invertibility of $\Sigma$ in the usual limiting regime.  The other three conditions prevent the existence of a node with a large degree (so-called hub) or a cluster of small hubs.  We show that all these conditions are satisfied if the graph is a $k$-MST, $k=O(1)$, based on the Euclidean distance for multivariate data.

\begin{theorem}\label{thm:asym-kMST}
When the graph is a $k$-MST, $k=O(1)$, based on the Euclidean distance, then $S \overset{\mathcal{D}}{\rightarrow} \chi_2^2$ in the usual limiting regime under the null hypothesis.

\end{theorem}

The proof of this theorem is in Appendix \ref{sec:proof-data}.  Since non-Euclidean data object can usually be embedded in a high-dimensional Euclidean space, this theorem is useful for object data as well when such a correspondence exists.

\subsection{Consistency results for multivariate data}
\label{sec:consistency}

\cite{henze1999multivariate} showed that the edge-count test on MST is consistent against all alternatives under the multivariate setting.  Extending their arguments, we can show that the new test statistic on $k$-MST, $k=O(1)$, is consistent against all alternatives under the multivariate setting.  

\begin{theorem}\label{thm:consistency}
For two continuous multivariate distributions, if the graph is a $k$-MST, $k=O(1)$, based on the Euclidean distance, the test based on $S$ is consistent against all alternatives in the usual limiting regime.
\end{theorem}

The complete proof of this theorem is in Appendix \ref{sec:proof-consistency}.  This consistency result is also useful for object data as many of them can be embedded in a high-dimensional Euclidean space.

\subsection{Accuracy of $p$-value approximations from the asymptotic null distribution for finite sample sizes}
\label{sec:accuracy}
The asymptotic distribution of the test statistic shown in Theorem \ref{thm:asym} can be used to calculate the approximate $p$-value of the test.  But how large a sample size must be so that the approximate $p$-value is good enough?  Here, we examine the approximate $p$-value for finite samples by comparing it to the permutation $p$-value calculated from 10,000 permutations, which serves as a good surrogate of the true permutation $p$-value.  Under different settings of sample sizes, we take the difference of the two $p$-values and see how close it is to 0.  

\begin{figure}[!htp]
$d=10:$\\
\includegraphics[width=\textwidth]{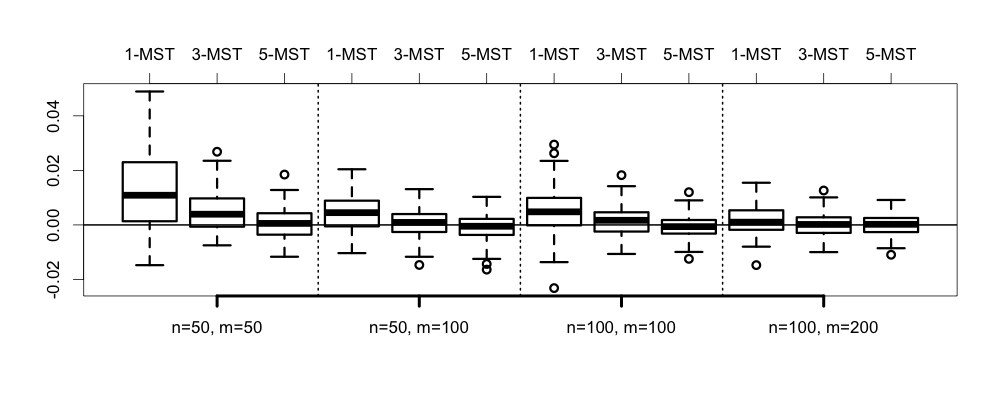}

$d=100:$\\
\includegraphics[width=\textwidth]{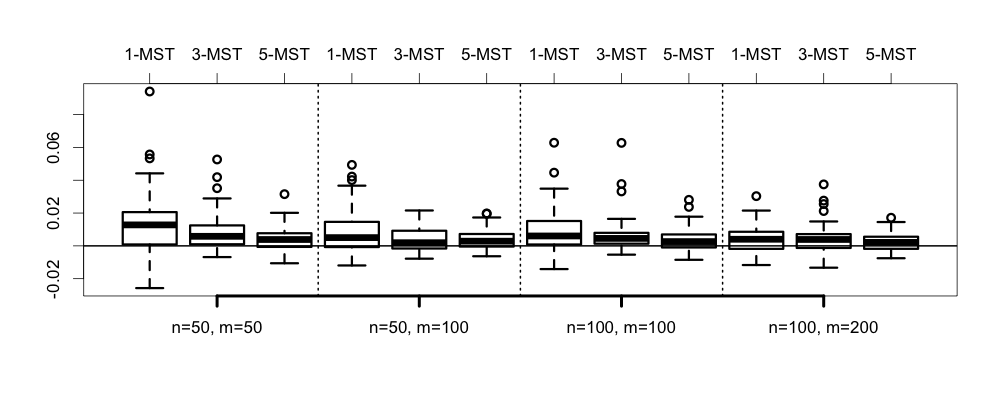}
\caption{Boxplots of the differences between the $p$-value based on the asymptotic distribution and the $p$-value calculated directly from 10,000 permutations (100 simulation runs for each setting. $F_\bX = F_\bY = \mathcal{N}(\mathbf{0}, I_d)$). }
\label{fig:pcheck}
\end{figure}

Figure \ref{fig:pcheck} shows boxplots of the differences of the two $p$-values (approximated $p$-value minus permutation $p$-value) from 100 simulation runs, under different choices of $n$, $m$, $d$ and the graph $G$.   We can see from the boxplots that the approximate $p$-value is slightly more conservative in general.  As the graph becomes denser, from a 1-MST to a 5-MST, the approximate $p$-value becomes more accurate, so the slightly denser graph is also preferred here.  The accuracy of the approximation increases as the sample sizes increases.  Increasing the dimension of the data slightly decreases the accuracy of the approximate $p$-value.  From the plots, sample sizes in hundreds are large enough to use the approximate $p$-value based on the asymptotic distribution.

\section{Real Data Examples}
\label{sec:application}
In this section, we illustrate the new test on two applications: The appraisal of covariate balance in a matched observational study, and the comparison of phone-call network data under two conditions.

\subsection{Covariates appraisal}

The new test is applied to a study on assessing a matched design for comparing ultimate educational attainment for students who start college at two-year vs. four-year colleges in the United States \citep{rouse1995democratization, heller2010using}.  In the study, 429 students starting at two-year colleges (the treatment group T) were matched to three nonoverlapping control groups of students attending four-year colleges (C-1, C-2, C-3) according to 20 observed covariates, including gender, ethnics, test score, etc.  Each matched control group contains 429 students.  The control groups are layered: the first control group (C-1) is an optimal pair matching; the second (C-2) is an optimal pair matching from the unused controls; the third (C-3) is an optimal pair matching from the still unused controls.   

The goal of the matching was to produce treated and control groups that had covariate balance, i.e., the same distribution of covariates, so it is important to appraise how well the matching is.  (See \cite{hansen2008covariate} for discussion of evaluating balance in matched observation studies.)  As there are 20 covariates in this case, it is not easy to appraise the matching through parametric approaches.  In \cite{heller2010using}, they appraised covariate balance by testing whether the distributions of covariates were the same in the treated and each control group (and also in each control group vs. each other control group) by using the MDP test.  Their results are shown in the first column ($R_0$: MDP) in Table \ref{tab:college} where the four groups (T, C-1, C-2, C-3) are compared two at a time with each other.  We also made the six comparisons through the edge-count test on MST ($R_0$: MST, the second column of Table \ref{tab:college}) and the new test on MST ($S:$ MST, the third column).  The same distance in \cite{heller2010using}, a ranked-based Mahalanobis distance, was used in constructing the MST.

From Table \ref{tab:college}, it is clear that C-3 is very different from the other three groups, so we focus on the comparisons among T, C-1 and C-2 (rows 1, 2 and 4 in the table).  In all three tests, the treatment group (T) is very similar to C-1, but significantly different from C-2.  The interesting part is the comparison between C-1 and C-2.  Both edge-count tests say that C-1 is not that different from C-2 (not rejected at 0.01 significance level), which is not completely different from but somewhat in opposition to the result that the treatment group is very different from C-2, given that T and C-1 are not close to being significantly different.  On the other hand, the results from the new test are much more consistent: The difference between the treatment group and C-2 and the difference between C-1 and C-2 are quite similar, which is in line with the result that the treatment group and C-1 are very similar.

\begin{table}
  \caption{\label{tab:college} $p$-values for comparing matched groups two at a time.}
  \centering
  \begin{tabular}{c|ccc}
\hline
    & \multicolumn{3}{c}{$p$-value} \\
Match & $R_0$: MDP & $R_0$: MST & $S$: MST \\ \hline
T versus C-1 & 0.66 & 0.91 & 0.20 \\ 
T versus C-2 & 0.00013 & 0.0020 & 0.0065 \\
T versus C-3 & $3.6\times 10^{-32}$ & $7.4\times 10^{-59}$ & $2.8\times 10^{-57}$ \\
C-1 versus C-2 & 0.028 & 0.010 & 0.0027 \\
C-1 versus C-3 & $1.3\times 10^{-25}$ & $2.5\times 10^{-48}$ & $8.1\times 10^{-48}$ \\
C2 versus C-3 & $1.2\times 10^{-17}$ & $7.5\times 10^{-27}$ & $1.9\times 10^{-25}$ \\ \hline
  \end{tabular}
\end{table}

\subsection{Social network}

The MIT Media Laboratory conducted a study following 106 subjects, students and staff in an institute, who used mobile phones with pre-installed software that can record call logs from July 2004 to June 2005 \citep{eagle2009inferring}.  
Given the richness of this data set, lots of aspects can be studied.  One question of interest is whether phone call patterns on \emph{weekdays} are different from those on \emph{weekends}.  They can be viewed as representations of professional relationship and personal relationship, respectively.
%
%

\begin{figure}[!htp]
\begin{center}
\includegraphics[width=0.5\textwidth]{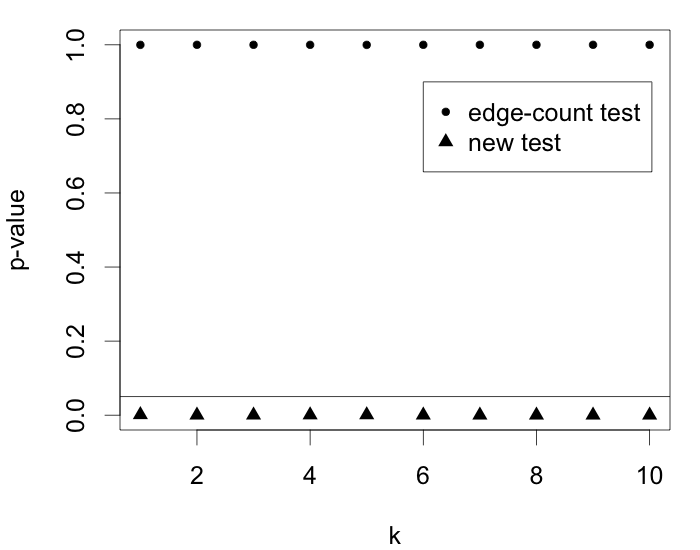}
\end{center}
\vspace{-2em}
\caption{The $p$-values of the edge-count tests (points) and the new tests (triangles) on $k$-MSTs with different $k$'s ($x$-axis).  The horizontal line is of level 0.05.}
\label{fig:pvalue}
\end{figure}

We bin the phone calls by day and, for each day, construct a directed phone-call network with the 106 subjects as nodes and a directed edge pointing from person $i$ to person $j$ if person $i$ made at least one call to person $j$ on that day. Among the 106 subjects, 87 of them made calls within themselves during the study. The distance between two networks is defined as the number of different directed edges in them (the direction matters).  $k$-MSTs are constructed on the pooled 330 networks based on this distance.  The $p$-values of the edge-count test and the new test on $k$-MSTs for different $k$'s are shown in Figure \ref{fig:pvalue}.  We see that the new test rejects the null hypothesis on all $k$-MSTs, from 1-MST to 10-MST, at 0.05 significance level, while the edge-count test does not reject on any of them.  So the conclusion from the new test is to reject the null hypothesis while the conclusion from the edge-count test is not to reject the null hypothesis.

Since the two tests provide contradictory conclusions, we next examine which one makes more sense.  Considering the 3-MST as an example, there are 535 between-sample edges, which is larger than its null expectation ($\bE(R_0)=403.3$).  According to the rationale of the edge-count test, the two samples are well connected, so they are from the same distribution.  
However, if we explore more into the 3-MST, we see that the phone-call networks on weekdays are much \emph{less} likely to be connected within themselves ($R_1=327$ compared to its null expectation $\bE(R_1)=504.2$), while the phone-call networks on weekends are much \emph{more} likely to be connected within themselves ($R_2=125$ compared to its null expectation $\bE(R_2)=79.5$).  Both are strong evidences indicating that the two samples are different.  The summary statistics are given in Table \ref{tab:network}.

\begin{table}[!htp]
\caption{\label{tab:network} Summary statistics for the 3-MST. Sample 1: phone-call networks on weekdays;  sample 2: phone-call networks on weekends.}

\centering
\begin{tabular}{|l|l|l|}
\hline
$R_0 = 535$ & $\bE(R_0)=403.3$ & $R_0-\bE(R_0)=131.7$ \\ \hline
$R_1 = 327$ & $\bE(R_1)=504.2$ & $R_1-\bE(R_1)=-177.2$ \\ \hline
$R_2 = 125$ & $\bE(R_2)=79.5$ & $R_2-\bE(R_2)=45.5$ \\ \hline
\end{tabular}
\end{table}

Hence, we see the same phenomenon here as that in moderate/high-dimensional data with a scale change: Both numbers of within-sample edges deviate from their null expectations, but the directions of the deviations are different.  As non-Euclidean data can usually be embedded into a high-dimensional Euclidean space, it is not surprising to see the same phenomenon in network data.  For this specific example, one plausible explanation is that personal relationship (reflected by call activities on weekends) is more stable than professional relationship (reflected by call activities on weekdays), thus the phone-call networks on weekends have a smaller ``variance" compared to those on weekdays.  

\section{Discussion}
\label{sec:discussion}

%

In this section, we briefly discuss several other test statistics along the same line as the new statistic $S$ by utilizing the deviation from the null expectation in both directions.  The following are four such test statistics:
\begin{align*}
T_1 & = |R_1-\mu_1| + |R_2-\mu_2| \\
T_2 & = \frac{|R_1-\mu_1|}{\sqrt{\Sigma_{11}}} + \frac{|R_2-\mu_2|}{\sqrt{\Sigma_{22}}} \\
T_3 & = (R_1-\mu_1)^2 + (R_2-\mu_2)^2 \\
T_4 & = \frac{(R_1-\mu_1)^2}{\Sigma_{11}} + \frac{(R_2-\mu_2)^2}{\Sigma_{22}}
\end{align*}

When $n=m$, $T_2$ is equivalent to $T_1$, and $T_4$ is equivalent to $T_3$.  When $n\neq m$, the performances of $T_2$ and $T_4$ are slightly better than those of $T_1$ and $T_3$ for location alternatives (see Tables \ref{tab:comp1} and \ref{tab:comp2}).

Comparing these four tests to the proposed test ($S$), we found that they all are comparable in low dimensions (Table \ref{tab:comp1}, $d=10$).  For data in high dimension (Table \ref{tab:comp2}, $d=100$), the proposed test ($S$) is much more powerful than these four tests ($T_1 - T_4$) for location-only alternatives, though the proposed test ($S$) is slightly less powerful than these four tests for scale-only alternatives.  Therefore, we still recommend the proposed test ($S$) in general scenarios unless one is very confident that the alternative is scale-only, under which $T_2$ or $T_4$ would be preferred.

\begin{table}[!htp]
\caption{\label{tab:comp1} Number of trials (out of 100) with significance less than 5\%, normal data, $d=10$.  The similarity graph is the MST based on the Euclidean distance.}
\centering
\begin{tabular}{c|ccccc}
\hline
\multicolumn{6}{c}{Location alternatives ($\Delta=1$)} \\ \hline
 & $T_1$ & $T_2$ & $T_3$ & $T_4$ & $S$ \\ \hline
$n=100, m=100$ & 33 & 33 & 29 & 29 & \textbf{36} \\ 
$n=100, m=200$ & 42 & 45 & 37 & 41 & \textbf{48} \\ \hline
\end{tabular}

\begin{tabular}{c|ccccc}
\hline
\multicolumn{6}{c}{Scale alternatives ($\sigma=1.1$)} \\ \hline
 & $T_1$ & $T_2$ & $T_3$ & $T_4$ & $S$ \\ \hline
$n=100, m=100$ & \textbf{45} & \textbf{45} & 42 & 42 & 37 \\ 
$n=100, m=200$ & \textbf{57} & 55 & 48 & 56 & 48 \\ \hline
\end{tabular}
\end{table}

\begin{table}[!htp]
\caption{\label{tab:comp2} Number of trials (out of 100) with significance less than 5\%, normal data, $d=100$.  The similarity graph is the MST based on the Euclidean distance.}
\centering

\begin{tabular}{c|ccccc}
\hline
\multicolumn{6}{c}{Location alternatives ($\Delta=2$)} \\ \hline
 & $T_1$ & $T_2$ & $T_3$ & $T_4$ & $S$ \\ \hline
$n=100, m=100$ & 20 & 20 & 28 & 28 & \textbf{71} \\ 
$n=100, m=200$ & 23 & 27 & 31 & 38 & \textbf{83} \\ \hline
\end{tabular}

\begin{tabular}{c|ccccc}
\hline
\multicolumn{6}{c}{Scale alternatives ($\sigma=1.05$)} \\ \hline
 & $T_1$ & $T_2$ & $T_3$ & $T_4$ & $S$ \\ \hline
$n=100, m=100$ & \textbf{84} & \textbf{84} & 82 & 82 & 71 \\ 
$n=100, m=200$ & \textbf{96} & 94 & 95 & 94 & 89 \\ \hline
\end{tabular}
\end{table}


\section{Conclusion}
\label{sec:conclusion}

We propose a new graph-based test statistic for comparing two distributions.  It utilizes a common pattern under the location alternatives and scale alternatives and has good power for detecting general alternatives for multivariate data and non-Euclidean data.  The asymptotic permutation null distribution of the test statistic is $\chi^2_2$ under some mild conditions on the graph. $P$-value approximation based on the asymptotic null distribution works well for samples in hundreds and beyond, making the test an easy off-the-shelf tool for analyzing large data sets.

 Under the multivariate setting, if the graph is a $k$-MST, $k=O(1)$, based on the Euclidean distance, then all the conditions on the graph for obtaining the asymptotic null distribution are satisfied and we have the unconditional limiting null distribution.  The test based on $k$-MST, $k=O(1)$, on Euclidean distance is also consistent against all alternatives.    

The new test has been applied to two real data sets.  In assessing the covariate balance in a matched observational study, the new test provides more consistent results than the existing graph-based tests.  In comparing network data under two conditions, the new test is able to capture the ``variance" difference in networks.

%
%

\section*{Acknowledgements}
\label{sec:acknowledgements}

Hao Chen is supported in part by NSF award DMS-1513653.  We thank Dylan Small for very helpful discussions and for kindly providing the data for the analysis of college students matchings.  We also thank two anonymous referees for very helpful comments.
 
\bibliographystyle{plainnat}
\bibliography{newtest}

\appendix

\section{Proofs}
\label{sec:proofs}

\subsection{Proof to Lemma \ref{lemma:EV}}
\label{sec:proof-lemma-EV}

Under permutation null distribution, we have
 \begin{align*}
    \bE R_1 & = \sum_{e\in G} \bP(J_e=1) = \sum_{(i,j)\in G} \bP(g_i=1, g_j=1) = |G| \frac{n(n-1)}{N(N-1)}. \\
    \bE (R_1^2) & = \sum_{e_1,e_2\in G} \bP(J_{e_1}=1, J_{e_2}=1) \\
       & = \sum_{(i,j)\in G} \bP(g_i=1, g_j=1) + \sum_{(i,j),(i,k)\in G;\ j\neq k} \bP(g_i=1, g_j=1, g_k=1) \\ 
       & \quad \quad + \sum_{(i,j),(k,l)\in G;\ i,j,k,l \text{ all different}} \bP(g_i=1, g_j=1, g_k=1, g_l=1) \\
       & = |G| \frac{n(n-1)}{N(N-1)} + 2C\frac{n(n-1)(n-2)}{N(N-1)(N-2)} \\
       & \quad + (|G|(|G|-1)-2C)\frac{n(n-1)(n-2)(n-3)}{N(N-1)(N-2)(N-3)}.
  \end{align*}
Then $\Sigma_{11}=\bE (R_1^2) - (\bE R_1)^2$ follows readily.  The expectation and variance of $R_2$ can be done in a similar manner.  For the covariance between $R_1$ and $R_2$, we have
\begin{align*}
  \bE(R_1R_2) & = \sum_{e_1,e_2\in G} \bP(J_{e_1}=1, J_{e_2}=2) \\
       & = \sum_{(i,j),(k,l)\in G;\ i,j,k,l \text{ all different}} \bP(g_i=1, g_j=1, g_k=2, g_l=2) \\
       & = (|G|(|G|-1)-2C)\frac{n(n-1)m(m-1)}{N(N-1)(N-2)(N-3)},
\end{align*}
and $\Sigma_{12}=\bE(R_1R_2)-\bE R_1 \bE R_2$ follows readily.

\subsection{Proof of Theorem \ref{thm:asym}}
\label{sec:proof-thm-asym}

The proof of Theorem \ref{thm:asym} relies on Stein's method.  Consider sums of the form $W=\sum_{i\in{\cal J}} \xi_i,$
where $\mathcal{J}$ is an index set and $\xi$ are random variables with $\bE \xi_i=0$, and $\bE (W^2)=1$.  The following assumption restricts the dependence between $\{\xi_i:~i \in \mathcal{J}\}$.
\begin{assumption} \cite[p.\, 17]{chen2005stein}
  \label{assump:LD}
For each $i\in{\cal J}$ there exists $K_i \subset L_i \subset {\cal J}$ such that $\xi_i$ is independent of $\xi_{K_i^c}$ and $\xi_{K_i}$ is independent of $\xi_{L_i^c}$.
\end{assumption}
We will use the following theorem in proving Theorem \ref{thm:asym}.
\begin{theorem}\label{thm:3.4} \cite[Theorem 3.4]{chen2005stein}
Under Assumption \ref{assump:LD}, we have
$$\sup_{h\in Lip(1)} |\bE h(W) - \bE h(Z)| \leq \delta,$$
where $Lip(1) = \{h: \mathbb{R}\rightarrow \mathbb{R} \}$, $Z$ has ${\cal N}(0,1)$ distribution and
 $$\delta = 2 \sum_{i\in{\cal J}} (\bE|\xi_i \eta_i\theta_i| + |\bE(\xi_i\eta_i)|\bE|\theta_i|) + \sum_{i\in{\cal J}} \bE|\xi_i\eta_i^2|$$
with $\eta_i = \sum_{j\in K_i}\xi_j$ and $\theta_i = \sum_{j\in L_i} \xi_j$, where $K_i$ and $L_i$ are defined in Assumption \ref{assump:LD}.
\end{theorem}

To prove Theorem \ref{thm:asym}, we take one step back to study the statistic under the bootstrap null distribution, which is defined as follows: For each observation, we assign it to be from sample $\bX$ with probability $n/N$, and from sample $\bY$ with probability $1-n/N$, independently of other observations.  Let $n_X$ be the number of observations that are assigned to be from sample $\bX$.  Then, conditioning on $n_X=n$, the bootstrap null distribution becomes the permutation null distribution.  We use $\bPB$, $\bEB$, $\bVB$ to denote the probability, expectation, and variance under the bootstrap null distribution, respectively.  (We here add the subscript $_{\mathbf{P}}$ to denote the corresponding quantities under the permutation null distribution.)

Let $p_n=n/N, q_n=1-p_n$, then $\lim_{N\rightarrow\infty} p_n = p, \lim_{N\rightarrow\infty}q_n=q$. Given that the $g_i$'s are independent under the bootstrap null distribution, we have
\begin{align*}
  \bEB R_1 & = |G|p_n^2 := \mu_1^B, \\
  \bEB R_2 & = |G|q_n^2 := \mu_2^B, \\
  \bVB(R_1) & = |G|p_n^2 q_n^2 + \sum_{i=1}^N|G_i|^2 p_n^3 q_n := (\sigma_1^B)^2, \\
  \bVB(R_2) & = |G|p_n^2 q_n^2 + \sum_{i=1}^N|G_i|^2 p_n q_n^3 := (\sigma_2^B)^2.
\end{align*}

Let 
\begin{align*}
  W_1^B & = \frac{R_1-\mu_1^B}{\sigma_1^B}, \quad W_1 = \frac{R_1-\mu_1}{\sigma_1}, \\
  W_2^B & = \frac{R_2-\mu_2^B}{\sigma_2^B}, \quad W_2 = \frac{R_2-\mu_2}{\sigma_2}, \\
  W_3^B & = \frac{n_X-n}{\sqrt{Np_n q_n}}.
\end{align*}

Under the conditions of Theorem \ref{thm:asym}, as $N\rightarrow\infty$, we can prove the following results:
\begin{enumerate}[(1)]
\item $(W_1^B, W_2^B, W_3^B)$ becomes multivariate Gaussian distributed under the bootstrap null.
\item $$\frac{\sigma_1^B}{\sigma_1}\rightarrow c_1, \quad \frac{\mu_1^B - \mu_1}{\sigma_1^B}\rightarrow 0; \quad \frac{\sigma_2^B}{\sigma_2}\rightarrow c_2, \quad \frac{\mu_2^B - \mu_2}{\sigma_2^B}\rightarrow 0,$$
where $c_1$ and $c_2$ are constants.
\item $|\lim_{N\rightarrow\infty}\bcorP(W_1,W_2)|<1$.
\end{enumerate}

From (1) and given that $\bVB(W_3^B)=1$, the conditional distribution of $(W_1^B, W_2^B)^\prime$ given $W_3^B$ is a bivariate Gaussion distribution under the bootstrap null distribution as $N\rightarrow \infty$.  Since the permutation null distribution is equivalent to the bootstrap null distribution given $W_3^B=0$, $(W_1^B,W_2^B)$ follows a bivariate Gaussian distribution under the permutation null distribution as $N\rightarrow\infty$.  Since 
$$ W_1 = \frac{\sigma_1^B}{\sigma_1}\left(W_1^B + \frac{\mu_1^B - \mu_1}{\sigma_1^B}\right), \quad  W_2 = \frac{\sigma_2^B}{\sigma_2}\left(W_2^B + \frac{\mu_2^B - \mu_2}{\sigma_2^B}\right),$$
given (2), we have $(W_1, W_2)$ follows a bivariate Gaussian distribution under the permutation null distribution as $N\rightarrow\infty$.  Together with (3), we have the conclusion in Theorem \ref{thm:asym}.  In the following, we prove the results (1)-(3).

To prove (1), by Cram$\acute{\text{e}}$r-Wold device, we only need to show that $W=a_1W_1^B + a_2W_2^B + a_3W_3^B$ is asymptotically Gaussian distributed for any combination of $a_1,a_2,a_3$ such that $\bVB(W)>0$.

Let 
\begin{align*}
  \xi_e & = a_1 \frac{I_{J_e=1} - p_n^2}{\sigma_1^B} + a_2 \frac{I_{J_e=2}-(1-p_n)^2}{\sigma_2^B}, \\
  \xi_i & = a_3 \frac{I_{g_i=0}-p_n}{\sqrt{Np_n(1-p_n)}}.
\end{align*}

Let $a=\max(|a_1|,|a_2|,|a_3|)$, $\sigma=\min(\sigma_1^B,\sigma_2^B,\sqrt{Np_n(1-p_n)})$, then $\sigma=O(N^{0.5})$, and $|\xi_e|\leq 2a/\sigma$, $|\xi_i|\leq a/\sigma$.  Let $\mathcal{J} = \{e\in G\} \cup \{1,\dots,N\}$.  

For $e=(e_-,e_+) \in \mathcal{J}$, let 
\begin{align*}
  K_e & = A_e\cup \{e_-,e_+\}, \\
  L_e & = B_e\cup\{\text{nodes in $A_e$}\}.
\end{align*}
Then $K_e$ and $L_e$ satisfy Assumption \ref{assump:LD}.  

For $i\in \{1,\dots,N\}$, let 
\begin{align*}
  K_i & = \{e\in G_i\}\cup \{i\}, \\
  L_i & = \{e\in G_{i,2}\}\cup \{\text{nodes in $G_i$}\}.
\end{align*}
Then $K_i$ and $L_i$ satisfy Assumption \ref{assump:LD}.

For $j\in \mathcal{J}$, let $\eta_j = \sum_{k\in K_j} \xi_k$, $\theta_j=\sum_{k\in L_j} \xi_k$.
By Theorem \ref{thm:3.4}, we have $\sup_{h\in Lip(1)} |\bEB h(W) - \bE h(Z)|\leq \delta$ for $Z\sim \mathcal{N}(0,1)$, where
\begin{align*}
  \delta & = \frac{1}{\sqrt{\bVB(W)}}\left(2\sum_{j\in\mathcal{J}}(\bEB|\xi_j\eta_j\theta_j| + |\bEB(\xi_j\eta_j)|\bEB|\theta_j|) + \sum_{j\in\mathcal{J}} \bEB|\xi_j\eta_j^2| \right) \\
  & \leq \frac{1}{\sqrt{\bVB(W)}} \left( 5 \sum_{e\in G} \frac{8a^3}{\sigma^3}(|A_e|+2)(|B_e|+|A_e|+1) + 5 \sum_{i=1}^N \frac{a^3}{\sigma^3}(|G_i|+1)(|G_{i,2}|+|G_i|+1) \right) \\
  & \leq \frac{1}{\sqrt{\bVB(W)}} \left( \frac{360 a^3}{\sigma^3} \sum_{e\in G} |A_e||B_e| + \frac{10 a^3}{\sigma^3} \sum_{i=1}^N (|G_i|+1)(|G_{i,2}|+1)\right).
\end{align*}

Notice that for $e=(i,j)$, we have $G_i, G_j\subseteq A_e$, $G_{i,2}, G_{j,2} \subseteq B_e$, so $(|G_i|+1)(|G_{i,2}|+1) \leq (|A_e|+1)(|B_e|+1)$.  For each node $i$, we can randomly pick an edge that has $i$ as one of its end points, then each edge in the graph can be picked at most twice since an edge only has two end points.  Therefore,
$$\sum_{i=1}^N (|G_i|+1)(|G_{i,2}|+1) \leq 2\sum_{e\in G}(|A_e|+1)(|B_e|+1) \leq 8 \sum_{e\in G}|A_e||B_e|. $$
Hence,
$$\delta \leq \frac{440 a^3}{\sqrt{\bVB(W)}} \frac{1}{\sigma^3}\sum_{e\in G} |A_e||B_e|.$$
Since $440 a^3/\sqrt{\bVB(W)}$ is of constant order, $\sigma=O(N^{0.5})$, when $\sum_{e\in G}|A_e||B_e|= o(N^{1.5})$, we have $\delta\rightarrow 0$ as $N\rightarrow\infty$.

Next we prove result (2).  Since $|G|=O(N)$, $\sum_{i=1}^N|G_i|^2-4|G|^2/N=O(N)$, let $\lim_{N\rightarrow\infty} |G|/N = b_1$ and $\lim_{N\rightarrow\infty} |G_i|^2/N-4|G|^2/N^2 = b_2$; $b_1,b_2\in(0,\infty)$.  Then $\lim_{N\rightarrow\infty} |G_i|^2/N = b_2+4b_1^2$, and
\begin{align*}
 \lim_{N\rightarrow \infty} \frac{\sigma_1^2}{N} & = p^2 q^2 (b_1 + b_2 p/q) = p^2 q^2 b_1 + p^3 q b_2
\end{align*}

%
so 
\begin{align*}
\lim_{N\rightarrow\infty} \frac{\sigma_1^B}{\sigma_1} = & \lim_{N\rightarrow\infty} \sqrt{\frac{ (|G|p_n^2 q_n^2 + \sum_i|G_i|^2 p_n^3 q_n)/N}{ \sigma_1^2/N}}  = \sqrt{1+\frac{4pb_1^2}{qb_1+pb_2}}.
\end{align*}
Similarly, we have
\begin{align*}
\lim_{N\rightarrow\infty} \frac{\sigma_2^B}{\sigma_2}  = \sqrt{1+\frac{4qb_1^2}{pb_1+qb_2}}.  
\end{align*}

Also,
\begin{align*}
  \mu_1^B-\mu_1 = |G|\frac{n^2}{N^2} - |G|\frac{n(n-1)}{N(N-1)} = -|G|\frac{nm}{N^2(N-1)},
\end{align*}
so
\begin{align*}
  \lim_{N\rightarrow\infty} \frac{\mu_1^B - \mu_1}{\sigma_1^B} & = - \lim_{N\rightarrow\infty} \frac{r(1-r)|G|/N}{\sigma_1^B} = 0,
\end{align*}
since $|G|=O(N), \sigma_1^B =O(N^{0.5})$.

Similarly, we have $$ \lim_{N\rightarrow\infty} \frac{\mu_2^B - \mu_2}{\sigma_2^B}=0.$$

Next, we prove result (3). We utilize the expression of $\Sigma$ in Remark \ref{remark:Sigma} and obtain
\begin{align*}
 \lim_{N\rightarrow\infty} & \bcorP(W_1,W_2)  =\lim_{N\rightarrow\infty} \frac{\bEP(R_1R_2)-\mu_1\mu_2}{\sigma_1\sigma_2} = \lim_{N\rightarrow\infty}\frac{\Sigma_{12}}{\sqrt{\Sigma_{11}\Sigma_{22}}}  \\
& = \frac{b_1-b_2}{\sqrt{(b_1+b_2p/q)(b_1+b_2q/p)}} = \frac{b_1-b_2}{\sqrt{(b_1-b_2)^2 + \frac{b_1b_2}{pq}}}.
\end{align*}

Since $\frac{b_1b_2}{pq}$ is strictly positive, we have $|\lim_{N\rightarrow\infty} \bcorP(W_1,W_2)|<1$.

\subsection{Proof of Theorem \ref{thm:asym-kMST}}
\label{sec:proof-data}

For the $k$-MST, we have $|G|=k(N-1)$. If $k=O(1)$, then $|G|=O(N)$.  Following the arguments in \cite{henze1999multivariate}, for $k$-MST constructed under Euclidean distance on an \emph{iid} sample, we have $\lim_{N\rightarrow\infty} \sum_{i=1}^N |G_i|^2/N =E(D_{d,k}^2)$ and $\sum_{e\in G}|A_e||B_e|/N =O(E(D_{d,k}^4))$, where $D_{d,k}$ is the degree of the vertex at the origin ($\mathbf{0}$) in the $k$-MST on a homogeneous Poisson process on $\mathbb{R}^d$ of rate 1, with a point added at the origin, and the expectation and variance are in terms of the Poisson process.  

Let $D_{d,(j)}$ be the degree of the origin in the $j$th MST.  Then $D_{d,k} = \sum_{j=1}^k D_{d,(j)}$.  Following Lemma 7 of \cite{aldous1992asymptotics}, $E(D_{d,(j)}) = 2$, so $E(D_{d,k}) = 2k$. Then
$$\lim_{N\rightarrow\infty}\frac{1}{N}\left(\sum_{i=1}^N |G_i|^2-\frac{4|G|^2}{N}\right) = E(D_{d,k}^2)-4k^2 = Var(D_{d,k}).$$

We only need to show that $Var(D_{d,k})$ and $E(D_{d,k}^4)$ are bounded.  Since $k=O(1)$, it is enough to show that $Var(D_{d,1})$ and $E(D_{d,1}^4)$ are bounded.   
 The part for $Var(D_{d,1})$ has been shown in \cite{henze1999multivariate}. Let $\alpha_{i,d} = P(D_{d,1}=i)$.  Then $E(D_{d,1}^4) = \sum_{i=1}^\infty i^4\alpha_{i,d}$.  When $d$ increases, it is more likely to have larger degrees (see Table 1 in \cite{henze1999multivariate}) and $E(D_{d,1}^4)$ becomes larger, so we only need to show that $E(D_{\infty,1}^4)$ is bounded. \cite{penrose1996random} showed that $\alpha_{i,d}\rightarrow \alpha_i$ as $d\rightarrow\infty$, where
$$\alpha_i = \int_0^1 \exp(-\psi(u))\frac{\psi(u)^{k-1}}{(k-1)!}du,\footnote{There is a typo in the equation for $\alpha_i$ in \cite{henze1999multivariate}, pp.~292.  See Proposition 2 in \cite{aldous1990random}.} \text{ and } \psi(u)=\int_0^u \frac{\log(1/x)}{1-x}dx, u<1.$$
According to these formulas, it is not hard to calculate the numerical value of $E(D_{\infty,1}^4)$, which is 63.3.  

\subsection{Proof of Theorem \ref{thm:consistency}}
\label{sec:proof-consistency}

Let the density functions of the two multivariate distributions be $f$ and $g$.  Following the approach in \cite{henze1999multivariate}, we have 
$$\frac{R_1}{N} \rightarrow k \int \frac{p^2f^2(x)}{pf(x)+qg(x)} dx \quad \text{almost surely, and}$$
$$\frac{R_2}{N}\rightarrow k \int \frac{q^2g^2(x)}{pf(x)+qg(x)} dx \quad \text{almost surely.}$$



Let $$\delta_1=\lim_{N\rightarrow\infty} \frac{R_1-\mu_1}{N}, \quad \delta_2=\lim_{N\rightarrow\infty} \frac{R_2-\mu_2}{N},$$
and $b=Var(D_{d,k})$ as defined in Appendix \ref{sec:proof-data}.  Then
\begin{align*}
\lim_{N\rightarrow\infty}\frac{S}{N} & = \lim_{N\rightarrow\infty}\left(\frac{R_1}{N} - \frac{\mu_1}{N}, \frac{R_2}{N} - \frac{\mu_2}{N}\right) \left(\frac{\Sigma}{N}\right)^{-1} \left(\begin{array}{c} \frac{R_1}{N} - \frac{\mu_1}{N} \\ \frac{R_2}{N} - \frac{\mu_2}{N} \end{array} \right) \\
& = \frac{1}{p^2q^2kb}(\delta_1,\delta_2)\left(\begin{array}{cc} pqk+q^2b & -pq(k-b) \\ -pq(k-b) & pqk+p^2b\end{array} \right)\left(\begin{array}{c} \delta_1 \\ \delta_2 \end{array} \right) \\
& = \frac{pqk(\delta_1-\delta_2)^2 + b(q\delta_1 +p\delta_2 )^2}{p^2q^2kb}.
\end{align*}

We next show that $q\delta_1 + p\delta_2 >0$ when $f$ and $g$ differ on a set of positive measure.  Noticing that
\begin{align*}
\frac{q\delta_1 + p\delta_2}{k} & = \int \frac{qp^2f^2(x) + pq^2g^2(x)}{pf(x)+qg(x)}dx - (qp^2+pq^2) = pq\left(\int\frac{p f^2(x) + q g^2(x)}{p f(x) + q g(x)} dx - 1 \right),
\end{align*}
since
\begin{align*}
 \int\frac{p f^2(x) + q g^2(x)}{p f(x) + q g(x)} dx - 1 & = \int\frac{pf(x)(f(x)-g(x))}{pf(x)+qg(x)} dx = \int\frac{qg(x)(g(x)-f(x))}{pf(x)+qg(x)} dx,
\end{align*}
we have
\begin{align*}
 \frac{q\delta_1 + p\delta_2}{kpq} & = q\int\frac{pf(x)(f(x)-g(x))}{pf(x)+qg(x)} dx + p\int\frac{qg(x)(g(x)-f(x))}{pf(x)+qg(x)} dx \\
& = pq\int\frac{(f(x)-g(x))^2}{pf(x)+qg(x)}dx.
\end{align*}
So $q\delta_1 + p\delta_2$ is strictly positive when $f$ and $g$ differ on a set of positive measure.

\end{document}